\documentclass[12pt,a4paper]{article}

\usepackage[DIV=13,BCOR=0mm]{typearea}

\usepackage{natbib}
\bibpunct[, ]{(}{)}{,}{a}{}{,}

\usepackage{wasysym}

\usepackage{graphicx}

\usepackage{hyperref}
\usepackage{hyperxmp}
\hypersetup{
    pdfauthor={Ralf Greve},
    pdftitle={MAIC-2 v2.3},
    pdfsubject={Planetary science},
    pdfkeywords={Mars, Planetary ice, Obliquity, Ice cap, %
                 Polar layered deposits, Ice age, Modelling},
    pdfcopyright={Creative Commons Attribution 4.0 International},
    pdflicenseurl={https://creativecommons.org/licenses/by/4.0/}}

\usepackage{multirow}

\usepackage{minibox}

\usepackage{lmodern} 

\begin{document}

\title{\textbf{MAIC-2 v2.3:\\%
       {\fontsize{19pt}{23pt}\selectfont{}A Fortran program for simulating
       the surface temperature, atmospheric water transport
       and surface glaciation of the planet Mars}}}

\author{\protect\rule{0mm}{3.5ex}Ralf GREVE$^{1,2}$,
        Bj\"orn GRIEGER$^{3}$,
        Oliver J. STENZEL$^{4}$}

\date{{\small%
    $^1$Institute of Low Temperature Science, Hokkaido University,
        Sapporo, Japan\\[0.5ex]%
    $^2$Arctic Research Center, Hokkaido University,
        Sapporo, Japan\\[0.5ex]%
    $^3$European Space Astronomy Centre (ESAC), Villanueva de la Ca\~{n}ada,
        Madrid, Spain\\[0.5ex]%
    $^4$Max Planck Institute for Solar System Research,
        G\"ottingen, Germany\\[2.5ex]%
    ORCID:\\[0.5ex]%
       RG: 0000-0002-1341-4777,
       BG: 0000-0002-9420-4295,
       OJS: 0000-0001-7253-8588\\[2.5ex]%
    Corresponding author: Ralf Greve (greve@lowtem.hokudai.ac.jp)}\\[2.5ex]%
    {\normalsize{}November 17, 2026}%
}

\maketitle

\begin{abstract}

The Mars Atmosphere-Ice Coupler \texttt{MAIC-2} is a computationally efficient, zonally-averaged model designed to simulate Martian surface temperature, atmospheric water transport and surface glaciation over timescales of \(10^5\) to \(10^7\) years. While sophisticated General Circulation Models (GCMs) provide high-resolution simulations of the Martian atmosphere, their high computational cost makes them largely unsuitable for studies covering millions of years of orbital evolution. \texttt{MAIC-2} addresses this gap by utilizing physically-based, zonally-averaged parameterizations, including a Local Insolation Temperature (LIT) scheme and a mass balance engine driven by free-convection and instantaneous mixing physics. Version 2.3 introduces a fully modularized Fortran architecture that improves maintainability and ensures that the software remains entirely self-contained, requiring no external numerical libraries or specialized data formats for execution. Demonstrated through a 20-million-year transient simulation, the model successfully reproduces the primary characteristics of the Martian polar layered deposits. \texttt{MAIC-2} provides the planetary science community with a versatile, open-source tool for investigating the long-term climatic impacts of orbital variations on the Martian surface.

\end{abstract}

\section{Introduction}
\label{sec-intro}

Understanding the history of the Martian climate and surface glaciation requires simulating processes over paleoclimatological timescales (\(10^5\) to \(10^7\) years). While sophisticated General Circulation Models (GCMs), such as the Mars Planetary Climate Model \citep[PCM;][]{forget_etal_1999, forget_etal_2022, millour_etal_2024} and the NASA Ames MGCM \citep{haberle_etal_2003, kahre_etal_2023, kahre_etal_2024}, provide high-resolution simulations of the Martian atmosphere for the present, other time-slices or short periods of time, they are computationally intensive because they solve the full equations of fluid dynamics and thermodynamics. This high computational cost makes them largely unsuitable for studies covering millions of years of orbital evolution.

\texttt{MAIC-2} (Mars Atmosphere-Ice Coupler Version 2) addresses this need by providing a computationally efficient model specifically designed for long-term paleoclimate studies. It simplifies the complex descriptions of the Martian climate found in GCMs by utilizing a set of physically-based, zonally-averaged parameterizations. This approach allows the model to bridge the gap between high-fidelity atmospheric simulations and the multi-million-year timescales required to investigate the evolution of surface ice and polar layered deposits. In Section~\ref{sec-maic}, we sketch the development history of the model, starting in the mid-2000s as a module of the 3D dynamic/thermo\-dynamic ice-sheet model \texttt{SICOPOLIS} \citep[SImulation COde for POLythermal Ice Sheets,][]{sico_git_v26_2026}, later evolving into a standalone program, and culminating in the current, thoroughly modernized version 2.3 with a fully modularized design. In Section~\ref{sec-simulation}, we discuss a previously unpublished simulation of the Martian glaciation from \(10\:\mathrm{Ma}\) ago until \(10\:\mathrm{Ma}\) into the future, driven directly by the zonally averaged insolation computed from time-dependent orbital parameters (obliquity, eccentricity, and solar longitude of perihelion) provided by a numerical model of the solar system \citep{laskar_etal_2004}. Section~\ref{sec-conclusion} concludes the paper.

\section{Mars Atmosphere-Ice Coupler}
\label{sec-maic}

\subsection{Legacy}
\label{ssec-maic-legacy}

The Mars Atmosphere-Ice Coupler \texttt{MAIC} came into existence as a module of the ice-sheet model \texttt{SICOPOLIS}. \texttt{MAIC} version 1.0, described and used in the studies by \citet{greve_etal_2004, greve_mahajan_2005}, consisted of parameterizations for the surface temperature and surface mass balance of the Martian polar layered deposits (PLDs) stongly influenced by approaches for terrestrial ice sheets. Pinning points for these parameterizations were a mean annual temperature at the north or south pole computed by a balance between incoming shortwave solar radiation and outgoing longwave thermal radiation, and an estimate of the present-day H\(_2\)O ice accumulation rate.

Version 1.5 \citep{greve_2008a} introduced the Local Insolation Temperature (LIT) scheme originally devised by B.~Grieger (personal communication 2004). The LIT scheme computes the surface temperature of the PLDs via a daily, latitude-dependent balance between incoming shortwave solar radiation and outgoing longwave thermal radiation, buffered by the formation of a seasonal CO\(_2\) ice cap that prevents the temperature from falling below the condensation temperature of CO\(_2\). As for the surface mass balance of H\(_2\)O ice, the equilibrium-line approach with the present-day accumulation rate as the main parameter already available in version 1.0 was retained.

In contrast to the previous versions, \texttt{MAIC} version 2, or \texttt{MAIC-2} for short, is a standalone program. Since it is not part of a dynamic/thermodynamic ice-sheet model anymore, glacial flow and englacial temperatures are not computed. Instead, \texttt{MAIC-2} assumes that the evolution of the PLDs is determined in good approximation solely by the surface mass balance. The physics of the model (as of version 2.1) is explained in the study by \citet{greve_etal_2010} and shall not be repeated here in detail. In version 2.2, an optional north-south gradient of the atmospheric water content that reflects the topographic asymmetry of the planet was added. This is still unpublished in the peer-reviewed literature, but it will be used here (see Section~\ref{ssec-set-up}). Key features and requirements fulfilled by \texttt{MAIC-2} include:

\begin{itemize}

\item\textbf{Computational efficiency:} 
\texttt{MAIC-2} allows for transient simulations over 10 million years or more, which is currently unfeasible for standard GCMs.

\item\textbf{Orbital forcing integration:} 
The model is driven directly by essential orbital parameters (obliquity, eccentricity, and solar longitude of perihelion), allowing researchers to study how shifts in Mars' orbit drive ``ice ages'' and the formation of PLDs.

\item\textbf{Coupled surface-atmosphere processes:} 
It integrates the LIT scheme with parameterizations for atmospheric water transport (via instantaneous mixing) and surface mass balance.

\item\textbf{Predictive accuracy for glaciation:} 
Despite its simplicity, the model produces a remarkably good agreement with observed present-day polar ice volumes and thicknesses, making it a useful (``All models are wrong, but some are useful''; George Box) tool for reconstructing past glacial distributions.

\item\textbf{Versatility in application:} 
\texttt{MAIC-2} can simulate both ``academic'' scenarios with constant orbital parameters and realistic transient histories, providing insights into the hemispheric asymmetry of ice distribution and the mobility of mid-latitude glaciers.

\end{itemize}

By providing an open-source, efficient alternative for simulating the Martian water cycle and surface ice evolution, \texttt{MAIC-2} enables the planetary science community to explore the long-term climatic impacts of orbital variations that shaped the Martian surface. Due to its simplicity, the model is also suitable for educational use.

\subsection{MAIC-2 v2.3}
\label{ssec-maic2-2p3}

The \texttt{MAIC-2} software is implemented in Fortran and has recently undergone a major architectural refactoring to transition from a collection of independent subroutines to a fully modularized design. This modernization, finalized in version 2.3 \citep[GitHub repository:][]{maic2_git_v2p3_2026}, encapsulates the physical parameterizations of the model within discrete Fortran modules, which improves code maintainability and ensures robust interface checking during compilation. By grouping related physical processes, such as the Local Insolation Temperature (LIT) scheme, atmospheric mixing, and surface mass balance, the structure of the program reflects now more clearly the coupled nature of the Martian climate system.

A primary design goal for \texttt{MAIC-2} is extreme ease of use and portability across different computing environments. The software is entirely self-contained and does not require any external tools, numerical libraries, or specialized data formats such as NetCDF or Make. By relying solely on standard Fortran features, the program eliminates the complex dependency chains often associated with climate models, making it exceptionally simple to install and execute on any system with a standard Fortran 95/2003 compiler. This lightweight approach ensures that researchers can quickly deploy the model without the overhead of managing external software environments.

The computational core of the program is built around a finite-difference/finite-volume discretization of the governing equations. The model operates on a one-dimensional latitudinal grid, typically using a \(1^\circ\) spacing to balance spatial resolution with the high computational efficiency required for long-term paleoclimate studies. The simulation is driven by an orbital driver that numerically integrates the relationship between solar longitude (\(L_\mathrm{s}\)) and time to account for the significant eccentricity of the Martian orbit. This ensures that the seasonal forcing (including the variations in daily mean insolation and the timing of the CO\(_2\) frost cycle) is physically accurate over timescales spanning millions of years.

\begin{figure}[ht]
\centering
\includegraphics[width=0.95\textwidth]{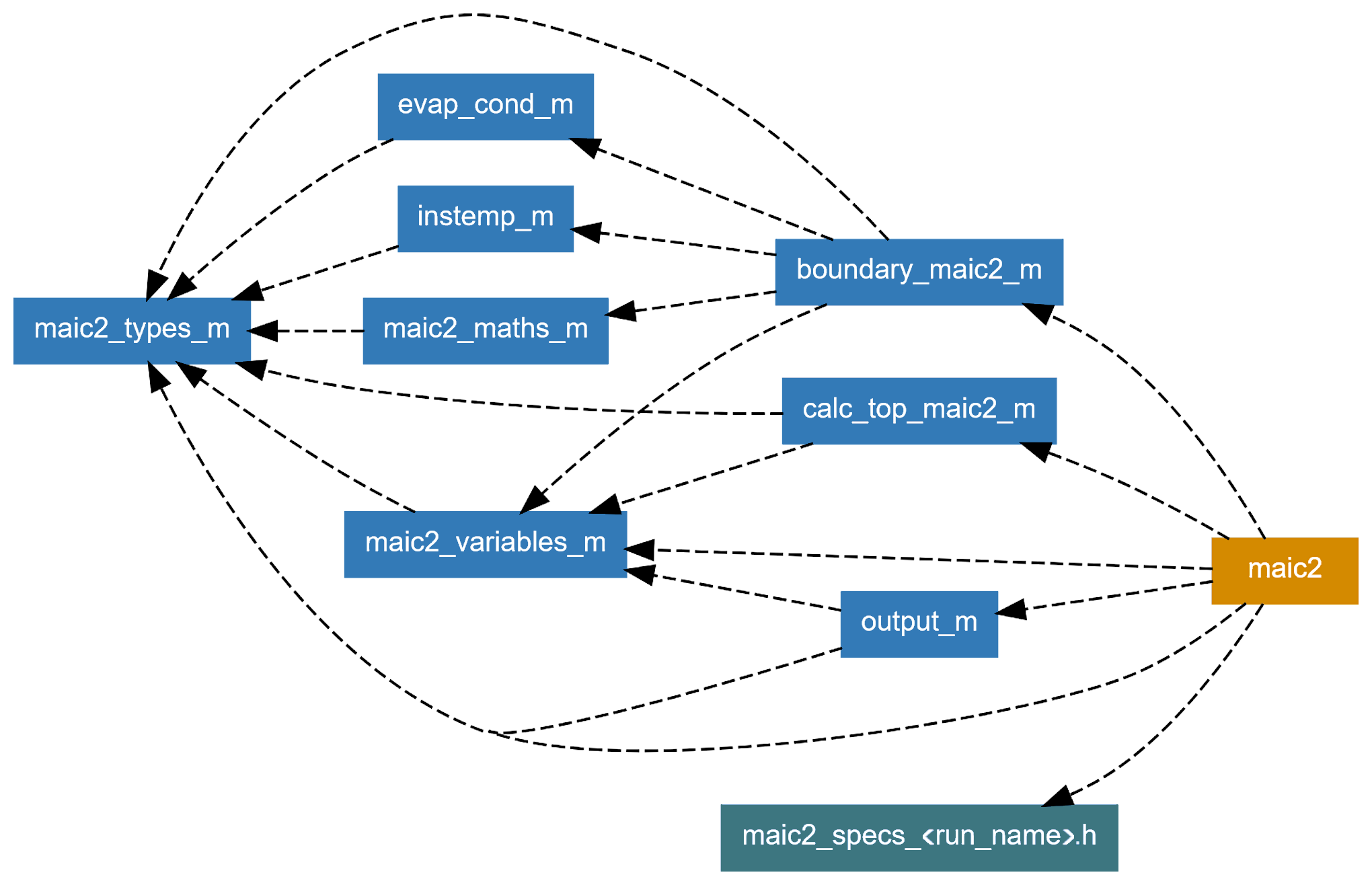}
\caption[Modular structure of \texttt{MAIC-2} v2.3.]{Modular structure of \texttt{MAIC-2} v2.3. All initializations are done in the main program \texttt{maic2}. This includes reading the physical parameters and the set-up from the run-specs header file \texttt{maic2\_specs\_<run\_name>.h}, where \texttt{<run\_name>} is a placeholder for the name of the simulation to be carried out. In the main loop (numerical forward-in-time integration), \texttt{maic2} carries out three steps: (1) Use the module \texttt{boundary\_maic2\_m} to update the orbital parameters, compute the surface temperature (LIT scheme, module \texttt{instemp\_m}) and the surface mass balance for H\(_2\)O ice (evaporation, atmospheric water transport and condensation, module \texttt{evap\_cond\_m}); (2) use the module \texttt{calc\_top\_maic2\_m} to update the topography (thickness) of the H\(_2\)O ice deposits; (3) use the module \texttt{output\_m} for data output in ASCII format. Auxiliary modules are \texttt{maic2\_types\_m} for declarations of kind types, \texttt{maic2\_variables\_m} for declarations of global variables, and \texttt{maic2\_maths\_m} for required mathematical tools.
\label{fig_maic2_structure}}
\end{figure}

The modular design specifically separates the calculation of surface temperatures from the mass balance engine. The LIT module determines temperatures by solving a radiation balance that accounts for the minimum threshold of CO\(_2\) condensation (\(148.7\:\mathrm{K}\) for a surface pressure of \(700\:\mathrm{Pa}\)) and the energy required for seasonal frost sublimation by making use of the fact that the same amount of heat that is released by CO\(_2\) condensation is required for complete sublimation. Simultaneously, the mass balance engine calculates water evaporation through free-convection physics and condensation via an instantaneous mixing approximation of atmospheric water vapor. These results are coupled to a static ice tracker that evolves the local ice thickness based on the net balance of these processes. The detailed structure of the program is shown in \autoref{fig_maic2_structure}. 

To support the scientific community, the \texttt{MAIC-2} package now includes a user manual in Markdown format, providing guidance on installation, set-up of simulations, compilation and execution, and output files. In addition, a developer manual in HTML format can be built by the tool FORD \citep[FORtran Documenter;][]{ford_git_v7p0p5_2024}. \autoref{fig_maic2_structure} is an extract from this developer manual.

Run-specs header files for all simulations discussed by \citet{greve_etal_2010}, as well as the two new simulations discussed below (Sect.~\ref{sec-simulation}), are contained in the \texttt{MAIC-2} GitHub repository. For a detailed list, see also the \texttt{MAIC-2} user manual. Note that the discontinuous numbering is due to intermediate test cases that are not discussed here or elsewhere for the sake of brevity.

As an aside, we have ported back the LIT module of \texttt{MAIC-2} v2.3 into the ice-sheet model \texttt{SICOPOLIS} \citep{sico_git_v26_2026}. This allows to carry out more refined simulations of the Martian PLDs that include englacial dynamics and thermodynamics.

\section{Paleoclimatic and future-climate simulation}
\label{sec-simulation}

\subsection{Orbital parameters}
\label{ssec-orbital-parameters}

\begin{figure}[ht]
\centering
\includegraphics[width=0.95\textwidth]{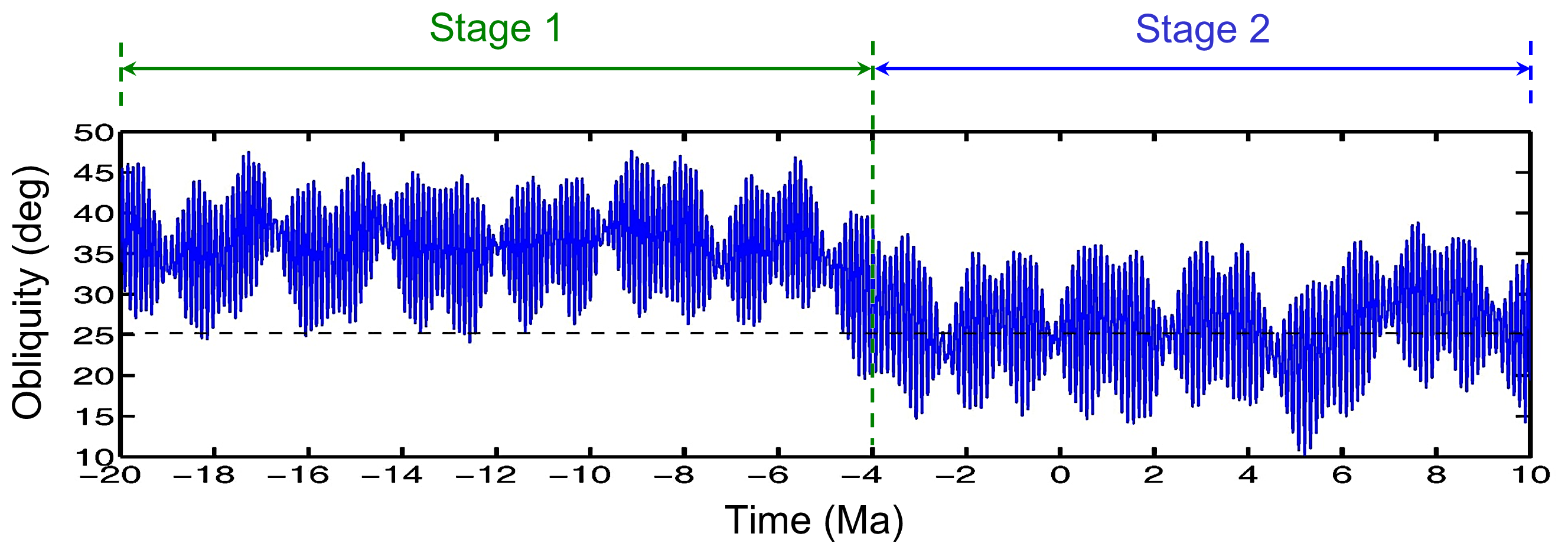}
\caption{Obliquity from \(20\:\mathrm{Ma}\) ago until \(10\:\mathrm{Ma}\) into the future \citep{laskar_etal_2004}.
\label{fig_obliquity}}
\end{figure}

Using a numerical model of the solar system, \citet{laskar_etal_2004} computed the evolution of the orbital parameters (obliquity, eccentricity, solar longitude of perihelion) of Mars. On time scales of 100 million years (\(100\:\mathrm{Ma}\)) and more, the solution is chaotic and therefore not predictable; however, a robust solution could be provided for the period from \(20\:\mathrm{Ma}\) ago until \(10\:\mathrm{Ma}\) into the future. The obliquity is shown in \autoref{fig_obliquity}. It features a main cycle with a period of \(125\:\mathrm{ka}\), a modulation with a period of \(1.3\:\mathrm{Ma}\), and a secular shift from \({\sim}35^\circ\) to \({\sim}25^\circ\) average obliquity at \({\sim}4\:\mathrm{Ma}\) ago, marked in the figure by ``Stage~1'' and ``Stage~2'', respectively.

\subsection{Set-up}
\label{ssec-set-up}

We illustrate the use of \texttt{MAIC-2} with the example of a transient simulation from \(10\:\mathrm{Ma}\) ago until \(10\:\mathrm{Ma}\) into the future, forced by the orbital parameters by \citet{laskar_etal_2004} as described above. A similar simulation from \(10\:\mathrm{Ma}\) ago until today was already discussed by \citet[][their simulation \#6]{greve_etal_2010}; cf.~\texttt{run\_t06} in the \texttt{MAIC-2} GitHub repository. It provided a good agreement between simulated and observed present-day volumes of the north-polar and south-polar layered deposits (NPLD, SPLD) and ice thicknesses at the respective poles. However, a weakness was that the simulation predicted that the present-day PLDs both formed during the last \(10\:\mathrm{Ma}\). This contradicts datings based on crater statistics which indicate that the basal unit of the NPLD and almost the entire SPLD are much older than a few millions of years \citep[e.g.,][]{herkenhoff_plaut_2000}.

\begin{figure}[ht]
\centering
\includegraphics[width=0.7\textwidth]{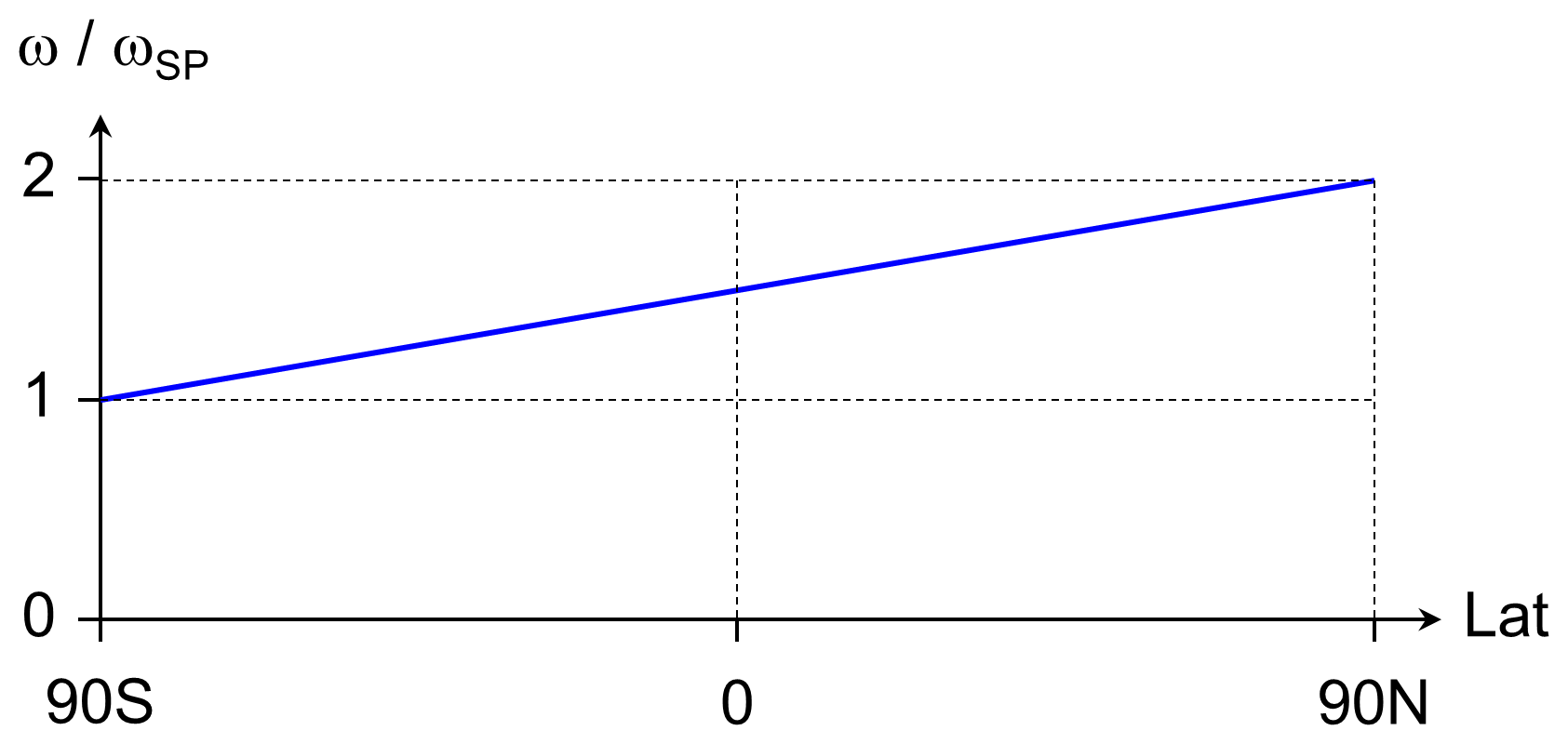}
\caption{Distribution of the atmospheric water content \(\omega\) relative to the value at the south pole, \(\omega_\mathrm{SP}\), with a prescribed north-south gradient. 
\label{fig_water_content}}
\end{figure}

Therefore, we employ a modified set-up. For the instantaneous mixing of atmospheric water vapour, we introduce a prescribed north-south gradient such that \(\omega_\mathrm{NP}/\omega_\mathrm{SP}=2\) (where \(\omega_\mathrm{NP}\) and \(\omega_\mathrm{SP}\) denote the depth-integrated water content \(\omega\) {[}mass per area{]} at the north and south pole, respectively). This choice is motivated by the topographically-forced asymmetry in the Martian circulation \citep{richardson_wilson_2002}. The resulting distribution of the water content is shown in \autoref{fig_water_content}.

Further changes of the set-up are as follows:

\begin{itemize}

\item Initial ice layer of \(5.25\:\mathrm{m}\) thickness on the entire surface {[}simulation \#6: \(19\:\mathrm{m}\){]}.

\item Exponent of the latitude-dependent parameterization for the daily cycle of the surface temperature: \(n=4\) {[}simulation \#6: \(n=3\); cf.~\citet[][Eq.~16]{greve_etal_2010}{]}.

\item Dust insulation factor \(E_0=0.05\) {[}simulation \#6: \(E_0=0.1\){]}.

\item Numerical time-step \(\Delta{}t=8\:\mathrm{sols}\) {[}simulation \#6: \(\Delta{}t=0.02\:\mathrm{a}\) (\(\approx{}7\:\mathrm{sols}\)){]}.

\end{itemize}

The simulation as specified above is referred to as simulation \#39 (from \(10\:\mathrm{Ma}\) ago until today) and simulation \#40 (continuation until \(10\:\mathrm{Ma}\) into the future). In the \texttt{MAIC-2} GitHub repository, the respective designations are \texttt{run\_t39} and \texttt{run\_t40}.

\subsection{Results}
\label{ssec-results}

\begin{figure}[ht]
\centering
\includegraphics[width=0.985\textwidth]{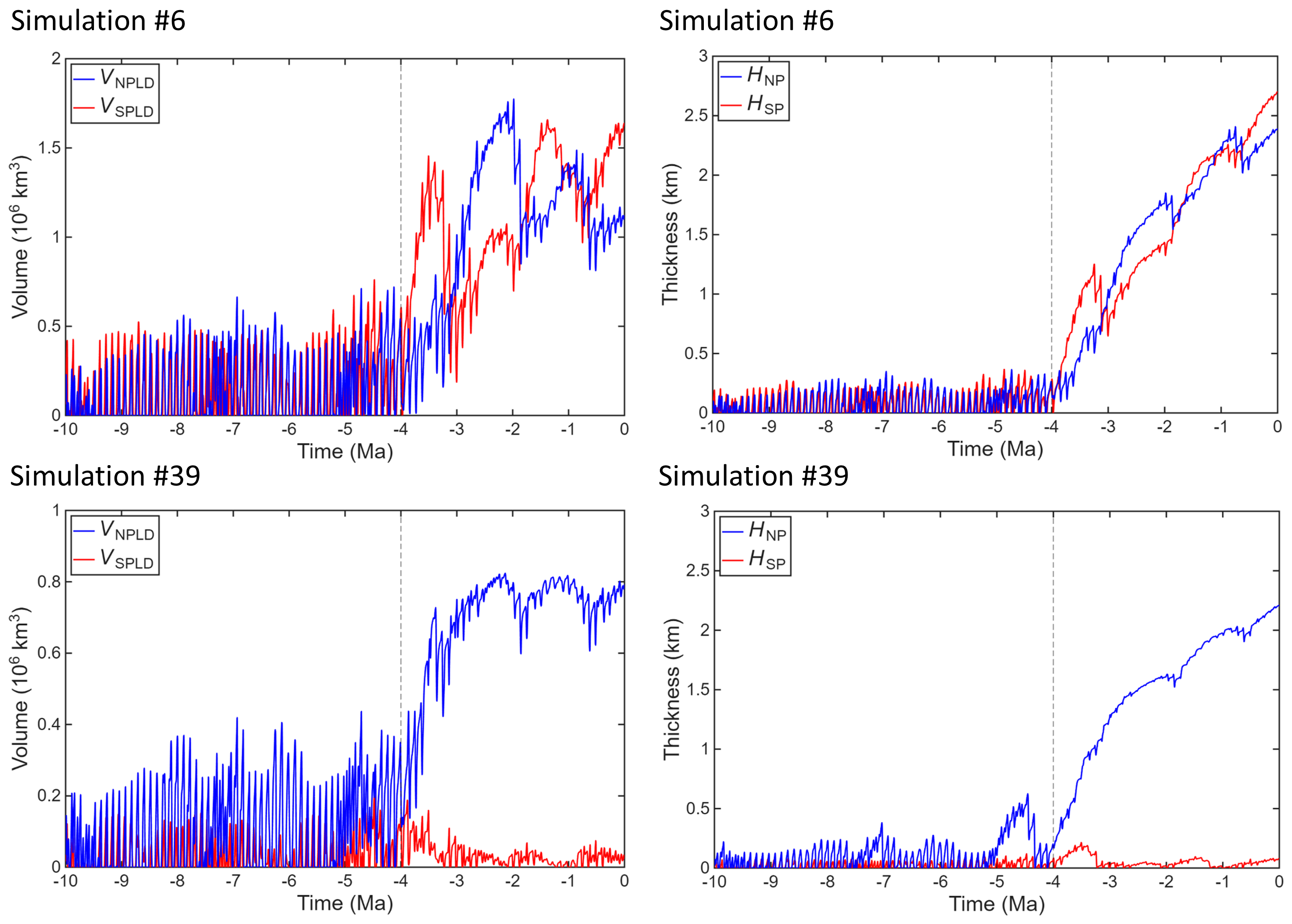}
\caption{Simulations \#6 (top panels) and \#39 (bottom panels): Volume of the north-polar and south-polar layered deposits (\(V_\mathrm{NPLD}\), \(V_\mathrm{SPLD}\); polewards of \(75^\circ\mathrm{N/S}\)), and ice thickness at the respective poles (\(H_\mathrm{NP}\), \(H_\mathrm{SP}\)). 
\label{fig_vol_thk_paleo}}
\end{figure}

\autoref{fig_vol_thk_paleo} depicts the volume of the PLDs (defined as the ice volume for latitudes \({\geq}75^\circ\mathrm{N}\) and \({\geq}75^\circ\mathrm{S}\), respectively) and ice thickness at the poles for our new simulation \#39 compared to simulation \#6 by \citet{greve_etal_2010}. In both cases, we see a very mobile glaciation in stage~1 (the period of high average obliquity prior to \(4\:\mathrm{Ma}\) ago), during which the ice is redistributed constantly between the high, mid and low latitudes, and ice thicknesses at any position on the planet never exceed some \(100\:\mathrm{m}\). During stage~2 (low average obliquity since \(4\:\mathrm{Ma}\) ago), massive PLDs form. As already mentioned above (Section~\ref{ssec-set-up}), the results of simulation \#6 contradict the greater age of the basal unit of the NPLD and almost the entire SPLD. By contrast, simulation \#39 produces only a massive glaciation in the north, while the southern glaciation remains very much limited.

Significantly, simulation \#39 predicts three large-scale erosional events at \({\sim}3.2\), \(1.9\) and \(0.7\:\mathrm{Ma}\) ago that interrupt relatively continuous growth of the NPLD in the past \(4\:\mathrm{Ma}\). This is, to first order, consistent with the finding of three periods of massive ablation within the internal stratigraphy of the NPLD \citep{holt_etal_2012, steel_holt_2012, smith_etal_2026}.

\begin{figure}[ht]
\centering
\includegraphics[width=0.71\textwidth]{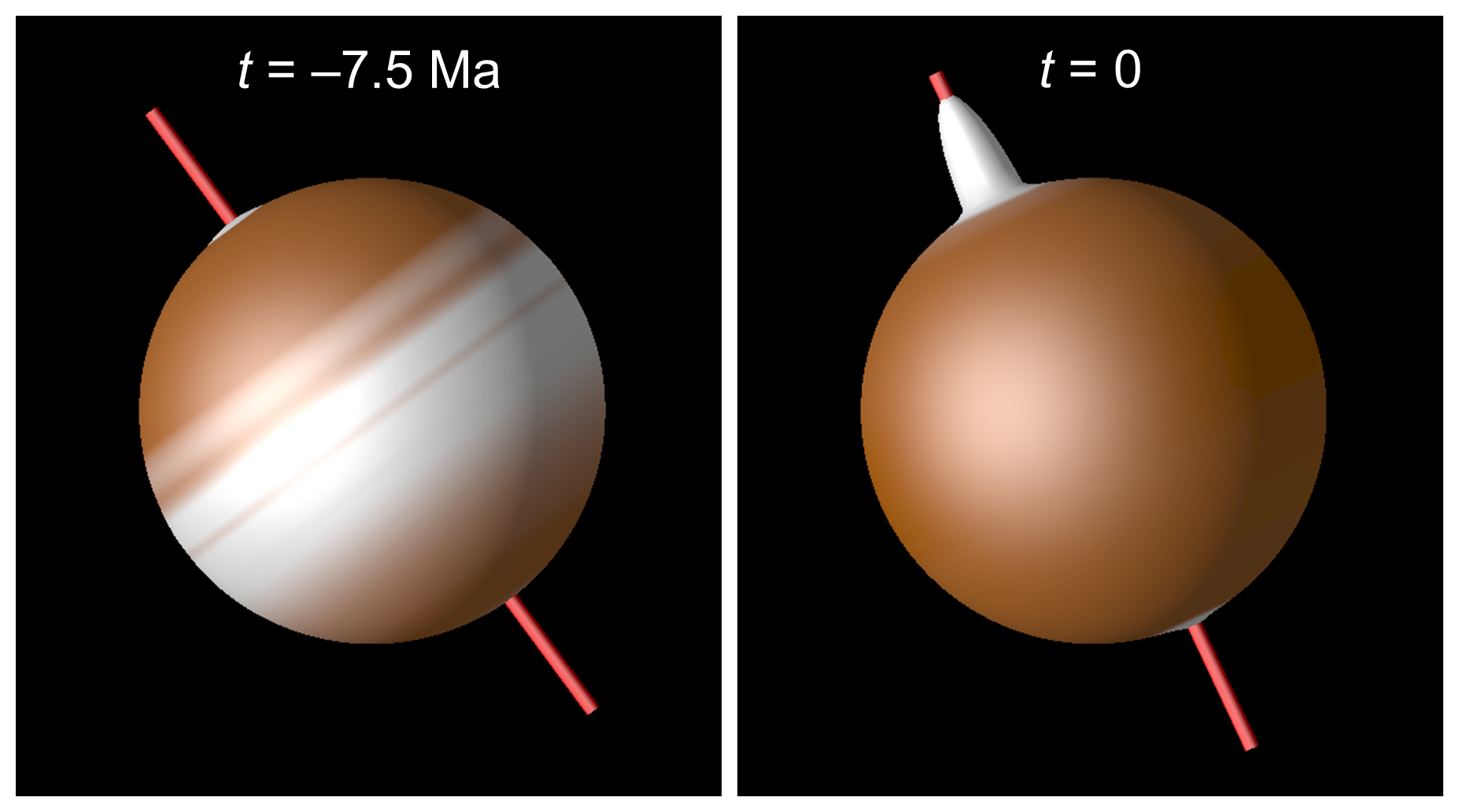}
\caption{Simulation \#39: Snapshots of the Martian glaciation for \(7.5\:\mathrm{Ma}\) ago (\(t=-7.5\:\mathrm{Ma}\), left) and today (\(t=0\), right). The ice thicknesses are exaggerated by a factor 750 relative to the size of the planet. (For a supplementary animation of the time-dependent glaciation, see https://doi.org/10.5281/zenodo.3700636.) 
\label{fig_thk_paleo_snap}}
\end{figure}

Snapshots of the Martian glaciation during stage~1 (\(7.5\:\mathrm{Ma}\) ago, obliquity \(36.4^\circ\)) and stage~2 (today, obliquity \(25.2^\circ\)) for simulation \#39 are shown in \autoref{fig_thk_paleo_snap}. The state at \(7.5\:\mathrm{Ma}\) ago is characterized by thin NPLD (thickness at the north pole \({\sim}60\:\mathrm{m}\)), almost no ice at the south pole, yet extensive ice cover in the mid and low latitudes. By contrast, the situation for today shows no ice outside the polar regions, NPLD with a volume of \(7.9\times{}10^5\:\mathrm{km}^3\) and a thickness at the pole of \(2.2\:\mathrm{km}\), and much smaller SPLD with a volume of \(2.3\times{}10^4\:\mathrm{km}^3\) and a thickness at the pole of merely \(77\:\mathrm{m}\).

To interpret these results, it must be noted that the basal unit of the NPLD and the bulk volume of the SPLD, which most likely predate the period of time considered here substantially (see above, Section~\ref{ssec-set-up}), are not part of the simulation. Considering this, the agreement between the simulated and observed recent deposits is quite good. The present-day NPLD excluding the basal unit has an estimated volume of \(8.2\times{}10^5\:\mathrm{km}^3\) \citep{putzig_etal_2009}, which matches our simulated figure very well. As for the SPLD, \citet{smith_etal_2016} report a volume of \(7\times{}10^3\:\mathrm{km}^3\) for its discontinuous, recent deposits. While this volume is a factor 3 smaller than our simulated counterpart, both numbers are small: the model successfully reproduces the fact that recent SPLD accumulation constitutes only a small part (of the order of 1\%) of the bulk volume of the SPLD.

\begin{figure}[ht]
\centering
\includegraphics[width=0.985\textwidth]{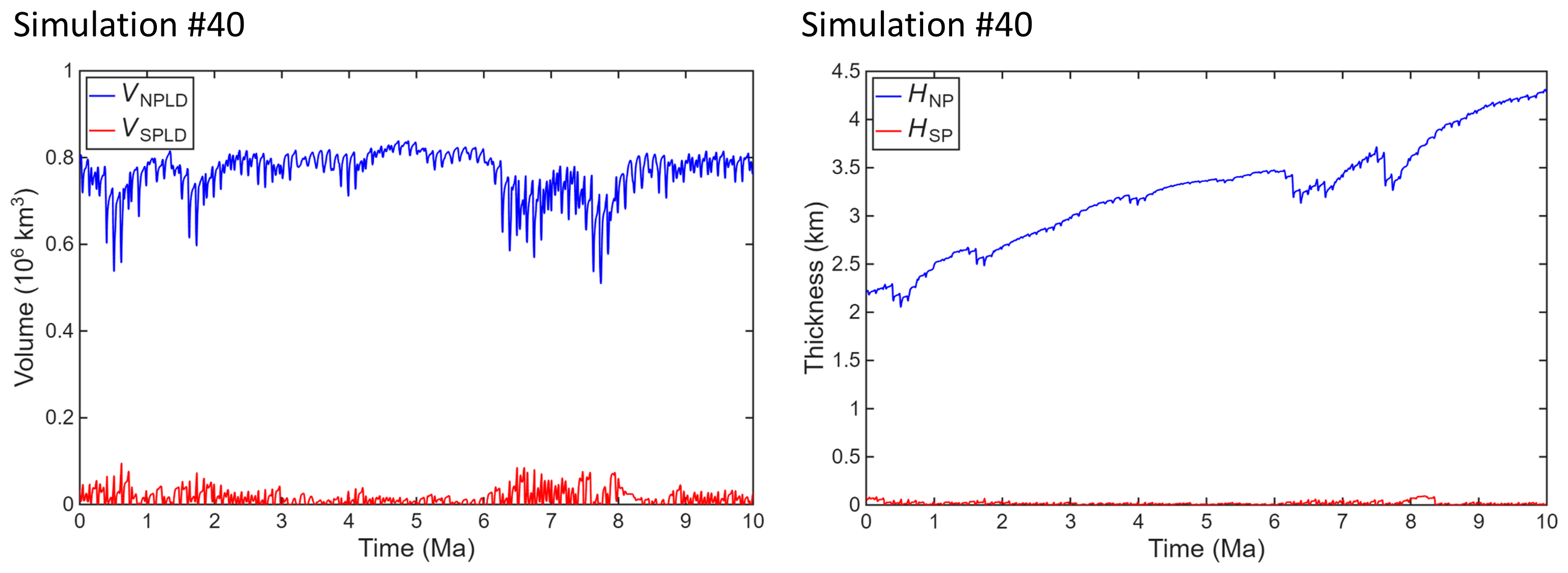}
\caption{Simulation \#40: Volume of the north-polar and south-polar layered deposits (\(V_\mathrm{NPLD}\), \(V_\mathrm{SPLD}\); polewards of \(75^\circ\mathrm{N/S}\)), and ice thickness at the respective poles (\(H_\mathrm{NP}\), \(H_\mathrm{SP}\)). 
\label{fig_vol_thk_future}}
\end{figure}

The results of simulation \#40, the continuation of simulation \#39 for \(10\:\mathrm{Ma}\) into the future, are shown in \autoref{fig_vol_thk_future}. Since this period is part of stage~2 (low average obliquity), no major regime shift is predicted. The volume of the NPLD remains large, with a generally increasing concentration of the ice close to the pole, towards a thickness of \({>}4\:\mathrm{km}\). By contrast, the volume of the SPLD remains small, with an ice thickness at the pole never exceeding \(100\:\mathrm{m}\), and episodically a complete loss of the ice.

\section{Conclusion}
\label{sec-conclusion}

We have introduced \texttt{MAIC-2 v2.3}, the latest version of the Mars Atmosphere-Ice Coupler that has a legacy reaching back to the mid-2000s. This updated version transitions the model into a fully modularized Fortran architecture, significantly enhancing the clarity and maintainability of the code. By prioritizing extreme ease of use and portability, the software operates as a self-contained tool that requires no external numerical libraries, allowing for quick deployment across diverse computing environments.

Our transient simulations demonstrate that, despite its computational efficiency, \texttt{MAIC-2} successfully reproduces the primary characteristics of Martian PLDs. Specifically, simulation \#39 shows strong agreement with observed present-day polar ice volumes and identifies large-scale erosional events consistent with internal NPLD stratigraphy. Furthermore, our forward-looking simulation \#40 suggests that under current low-obliquity conditions, the NPLD will continue to grow toward thicknesses exceeding \(4\:\mathrm{km}\). 

Ultimately, \texttt{MAIC-2 v2.3} provides the planetary science community with a versatile, open-source platform for investigating the long-term climatic impacts of orbital variations on the Martian surface. Beyond its research applications, the lightweight design and physical transparency of the model make it an excellent tool for educational use in planetary science and climate modelling curricula.

\section*{Acknowledgements}

We thank Isaac B. Smith (York University, Toronto, Ontario, Canada; Planetary Science Institute, Lakewood, Colorado, USA) for his interest in the model results, which encouraged us to resume development of \texttt{MAIC-2} after a long hiatus.

We acknowledge financial support from the German Research Foundation (Deutsche Forschungsgemeinschaft, DFG) via the Priority Programme ``Mars and the Terrestrial Planets'', and from the Institute of Low Temperature Science, Hokkaido University, via a Leadership Research Grant.

\section*{Author contributions} 

Ralf Greve conceptualized this study. All authors developed the MAIC-2 code. Ralf Greve carried out the simulations. Ralf Greve and Bj\"orn Grieger created the figures. Bj\"orn Grieger created the supplementary animation. All authors discussed and interpreted the results. Ralf Greve wrote the manuscript with contributions from all authors.

\section*{Open science}

MAIC-2 is free and open-source software, published in a persistent Git repository hosted by GitHub \citep{maic2_git_v2p3_2026}. The tagged version 2.3, as well as the output data produced for this study, are also available on Zenodo (https://doi.org/10.5281/zenodo.3696080).

\section*{Competing interests} 

There are no conflicts of interest to report, financial or otherwise.


\bibliographystyle{ralf}
\bibliography{references}

\end{document}